\documentclass{sig-alternate-05-2015}
\usepackage{times,amsmath,amsfonts,amssymb,epstopdf,xspace}
\usepackage{graphicx}
\usepackage{hyperref}
\usepackage[usenames]{color}

\newcommand{\eqn}[1] {(\ref{eqn:#1})}
\makeatletter
\newcommand{\pushright}[1]{\ifmeasuring@#1\else\omit\hfill$\displaystyle#1$\fi\ignorespaces}
\makeatother

\newcommand{\gadget}{{\sc Gadget-2}\xspace}
\newcommand{\swift}{{\sc swift}\xspace}
\newcommand{\qs}{{\sc QuickSched}\xspace}

\newcommand{\web}{\url{www.swiftsim.com}}

\begin{document}

\CopyrightYear{2016}
\setcopyright{acmlicensed}
\conferenceinfo{PASC '16,}{June 08 - 10, 2016, Lausanne, Switzerland}
\isbn{978-1-4503-4126-4/16/06}\acmPrice{\$15.00}
\doi{http://dx.doi.org/10.1145/2929908.2929916}

\title{{\ttlit SWIFT}: Using task-based parallelism, fully asynchronous
communication, and graph partition-based domain decomposition for
strong scaling on more than 100\,000 cores.}

\numberofauthors{4}
  
\author{
\alignauthor
       Matthieu~Schaller\\
       \affaddr{Institute for Computational Cosmology (ICC)}\\
       \affaddr{Department of Physics}\\
       \affaddr{Durham University}\\
       \affaddr{Durham DH1 3LE, UK}\\
       \email{\footnotesize \url{matthieu.schaller@durham.ac.uk}}
       \alignauthor
      Pedro~Gonnet\\
      \affaddr{School of Engineering and Computing Sciences}\\
      \affaddr{Durham University}\\
      \affaddr{Durham DH1 3LE, UK}\\ \vspace{1ex}
      \affaddr{Google Switzerland GmbH}\\
      \affaddr{8002 Z\"urich, Switzerland}\\
\and
\alignauthor
       Aidan~B.~G.~Chalk\\
       \affaddr{School of Engineering and Computing Sciences}\\
       \affaddr{Durham University}\\
       \affaddr{Durham DH1 3LE, UK}\\
\alignauthor
       Peter~W.~Draper\\
       \affaddr{Institute for Computational Cosmology (ICC)}\\
       \affaddr{Department of Physics}\\
       \affaddr{Durham University}\\
       \affaddr{Durham DH1 3LE, UK}\\
}

\date{\today}

\maketitle

%#####################################################################################################

\begin{abstract}
  We present a new open-source cosmological code, called \swift, designed to
  solve the equations of hydrodynamics using a particle-based approach (Smooth
  Particle Hydrodynamics) on hybrid shared / distributed-memory architectures.
  \swift was designed from the bottom up to provide excellent {\em strong scaling}
  on both commodity clusters (Tier-2 systems) and Top100-supercomputers
  (Tier-0 systems), without relying on architecture-specific features
  or specialized accelerator hardware.
  This performance is due to three main computational approaches:

  \begin{itemize}
    
    \item \textbf{Task-based parallelism} for shared-memory
      parallelism, which provides fine-grained load balancing and
      thus strong scaling on large numbers of cores.

    \item \textbf{Graph-based domain decomposition}, which uses
      the task graph to decompose the simulation
      domain such that the {\em work}, as opposed to just the {\em data},
      as is the case with most partitioning schemes, is equally distributed
      across all nodes.

    \item \textbf{Fully dynamic and asynchronous communication},
      in which communication is modelled as just another task in
      the task-based scheme, sending data whenever it is ready and
      deferring on tasks that rely on data from other nodes
      until it arrives.

  \end{itemize}
  
  In order to use these approaches, the code had to be re-written from
  scratch, and the algorithms therein adapted to the task-based paradigm.
  As a result, we can show upwards of 60\% parallel efficiency for 
  moderate-sized problems when increasing the number of cores 512-fold,
  on both x86-based and Power8-based architectures.
    
\end{abstract}


%#####################################################################################################

\section{Introduction}

For the past decade physical limitations have kept the speed of
individual processor cores constrained, so instead of getting {\em
  faster}, computers are getting {\em more parallel}.  Systems
containing up to 64 general-purpose cores are becoming commonplace,
and the number of cores can be expected to continue growing
exponentially, e.g.~following Moore's law, much in the same way
processor speeds were up until a few years ago.

As a consequence, in order to get faster, computer programs need to
rely on better exploitation of this massive parallelism, i.e.~ they
need to exhibit {\em strong scaling}, the ability to run roughly $N$
times faster when executed on $N$ times as many processors.  Without
strong scaling, massive parallelism can still be used to tackle ever
{\em larger} problems, but not make fixed-size problems {\em faster}.

Although this switch from growth in speed to growth in parallelism
has been anticipated and observed for quite some time, very little
has changed in terms of how we design and implement parallel
computations.
Branch-and-bound synchronous parallelism using
OpenMP\cite{ref:Dagum1998} and MPI\cite{ref:Snir1998}, as well as domain
decompositions based on geometry or space-filling curves \cite{warren1993parallel}
are still commonplace, despite both the
architectures and problem scales having changed dramatically since
their introduction.

The design and implementation of \swift\footnote{
\swift is an open-source software project and the latest version of
the source code, along with all the data needed to run the test cased
presented in this paper, can be downloaded at \web.}
\cite{gonnet2013swift,theuns2015swift,ref:Gonnet2015}, a large-scale
cosmological simulation code built from scratch, provided the perfect
opportunity to test some newer
approaches, i.e.~task-based parallelism, fully asynchronous communication, and
graph partition-based domain decompositions.

This paper describes these techniques, which are not exclusive to
cosmological simulations or any specific architecture, as well as
the results obtained with them.
While \cite{gonnet2013swift,ref:Gonnet2015} already describes the underlying algorithms
in more detail, in this paper we focus on the parallelization strategy
and the results obtained  on larger Tier-0 systems.

%#####################################################################################################

\section{Particle-based hydrodynamics}

Smoothed Particle Hydrodynamics \cite{Gingold1977,Price2012} (SPH) uses
particles to represent fluids.  Each particle $p_i$ has a position $\mathbf
x_i$, velocity $\mathbf v_i$, internal energy $u_i$, mass $m_i$, and a smoothing
length $h_i$.  The particles are used to interpolate any quantity $Q$ at any
point in space as a weighted sum over the particles:
\begin{equation}
    Q(\mathbf r) = \sum_i m_i \frac{Q_i}{\rho_i} W( \|\mathbf r - \mathbf r_i\| , h )
    \label{eqn:interp}
\end{equation}
where $Q_i$ is the quantity at the $i$th particle, $h$ is the {\em smoothing
  length}, i.e.~the radius of the sphere within which data will be considered
for the interpolation, and $W(r,h)$ is the {\em smoothing kernel} or {\em
  smoothing function}.  Several different forms for $W(r,h)$ exist, each with
their own specific benefits and drawbacks.

The particle density $\rho_i$ used in \eqn{interp} is itself computed similarly:
\begin{equation}
    \rho_i = \sum_{j,~r_{ij} < h_i} m_j W(r_{ij},h_i)
    \label{eqn:rho}
\end{equation}
where $r_{ij} = \|\mathbf{r_i}-\mathbf{r_j}\|$ is the Euclidean distance between
particles $p_i$ and $p_j$.  In compressible simulations, the smoothing length
$h_i$ of each particle is chosen such that the number of neighbours with which
it interacts is kept more or less constant, and can result in smoothing lengths
spanning several orders of magnitudes within the same simulation.

Once the densities $\rho_i$ have been computed, the time derivatives of the
velocity and internal energy, which require $\rho_i$, are
computed as follows:
\begin{align}
    \frac{dv_i}{dt} & = -\sum_{j,~r_{ij} < \hat{h}_{ij}} m_j \left[
        \frac{P_i}{\Omega_i\rho_i^2}\nabla_rW(r_{ij},h_i)\right. + \label{eqn:dvdt}\\
        & \pushright{\left.\frac{P_j}{\Omega_j\rho_j^2}\nabla_rW(r_{ij},h_j) \right], \nonumber} \\ 
    \frac{du_i}{dt} & = \frac{P_i}{\Omega_i\rho_i^2} \sum_{j,~r_{ij} < h_i} m_j(\mathbf v_i - \mathbf v_j) \cdot \nabla_rW(r_{ij},h_i), \label{eqn:dudt}
\end{align}
where $\hat{h}_{ij} = \max\{h_i,h_j\}$, and the particle pressure $P_i=\rho_i
u_i (\gamma-1)$ and correction term $\Omega_i=1 +
\frac{h_i}{3\rho_i}\frac{\partial \rho}{\partial h}$ are computed on the fly.

The computations in \eqn{rho}, \eqn{dvdt}, and \eqn{dudt} involve finding all
pairs of particles within range of each other.  Any particle $p_j$ is {\em
  within range} of a particle $p_i$ if the distance between $p_i$ and $p_j$ is
smaller or equal to the smoothing distance $h_i$ of $p_i$, e.g.~as is done in
\eqn{rho}.  Note that since particle smoothing lengths may vary between
particles, this association is not symmetric, i.e.~$p_j$ may be in range of
$p_i$, but $p_i$ not in range of $p_j$.  If $r_{ij} < \max\{h_i,h_j\}$, as is
required in \eqn{dvdt}, then particles $p_i$ and $p_j$ are within range {\em of
each other}.

The computation thus proceeds in two distinct stages that are evaluated
separately:
\begin{enumerate}
    \item {\em Density} computation: For each particle $p_i$,
        loop over all particles $p_j$ within range of $p_i$ and evaluate
        \eqn{rho}.
    \item {\em Force} computation: For each particle $p_i$,
        loop over all particles $p_j$
        within range of each other and evaluate \eqn{dvdt} and \eqn{dudt}.
\end{enumerate}

Finding the interacting neighbours for each particle constitutes the
bulk of the computation.  Many codes, e.g. in Astrophysics simulations
\cite{Gingold1977}, rely on spatial {\em trees} for neighbour finding
\cite{Gingold1977,Hernquist1989,Springel2005,Wadsley2004}, i.e.~$k$-d
trees \cite{Bentley1975} or octrees \cite{Meagher1982} are used to
decompose the simulation space.  In Astrophysics in particular,
spatial trees are also a somewhat natural choice as they are used to
compute long-range gravitational interactions via a Barnes-Hut
\cite{Barnes1986} or Fast Multipole \cite{Carrier1988} method.  The
particle interactions are then computed by traversing the list of
particles and searching for their neighbours in the tree.  In current
state-of-the-art simulations (e.g. \cite{Schaye2015}), the gravity
calculation corresponds to roughly one quarter of the calculation time
whilst the hydrodynamics scheme takes approximately half of the
total time. The remainder is spent in the astrophysics modules, which
contain interactions of the same kind as the SPH sums presented
here. Gravity could, however, be vastly sped-up compared to commonly
used software by employing more modern algorithms such as the Fast
Multipole Method \cite{Carrier1988}.

Although such tree traversals are trivial to parallelize, they
have several disadvantages, e.g.~with regards to computational
efficiency, cache efficiency, and exploiting symmetries in the
computation (see \cite{ref:Gonnet2015} for a more detailed
analysis).

%#####################################################################################################

\section{Parallelization strategy}

One of the main concerns when developing \swift was to break
with the branch-and-bound type parallelism inherent to parallel
codes using OpenMP and MPI, and the constant synchronization
between computational steps it results in.

If {\em synchronization} is the main problem, then {\em
  asynchronicity} is the obvious solution.  We therefore opted for a
{\em task-based} approach for maximum single-node, or shared-memory,
performance.  This approach not only provides excellent load-balancing
on a single node, it also provides a powerful model of the computation
that can be used to distribute the work equitably over a set of
distributed-memory nodes using general-purpose graph partitioning
algorithms.  Finally, the necessary communication between nodes can
itself be modelled in a task-based way, interleaving communication
seamlessly with the rest of the computation.

\subsection{Task-based parallelism}

Task-based parallelism is a shared-memory parallel programming
paradigm in which a computation is broken-down in to a set of
{\em tasks} which can be executed  concurrently.
In order to ensure that the tasks are executed in the right
order, e.g.~that data needed by one task is only used once it
has been produced by another task, {\em dependencies} between
tasks are specified and strictly enforced by a task scheduler.
Additionally, if two tasks require exclusive access to the same
resource, yet in no particular order, they are treated as
{\em conflicts} and the scheduler ensures that they are not executed
concurrently.
Computations described in this way then parallelize trivially:
each processor repeatedly grabs a task for which all dependencies
have been satisfied and executes it until there are no tasks left.

The main advantages of using a task-based approach are
\begin{itemize}
    \item The order in which the tasks are processed, and how they
        are assigned to each processor is completely
        dynamic and adapts automatically to load imbalances.
    \item If the dependencies and conflicts are specified correctly,
        there is no need for expensive explicit locking, synchronization,
        or atomic operations to deal with most concurrency problems.
    \item Each task has exclusive access to the data it is working on,
        thus improving cache locality and efficiency.
\end{itemize}
Task-based parallelism is not a particularly new concept and therefore
several implementations thereof exist, e.g.~Cilk \cite{ref:Blumofe1995},
QUARK \cite{ref:QUARK}, StarPU \cite{ref:Augonnet2011},
SMP~Superscalar \cite{ref:SMPSuperscalar}, OpenMP~3.0 \cite{ref:Duran2009},
Intel's TBB \cite{ref:Reinders2007}, and, to some extent,
Charm++ \cite{ref:Kale1993}.

Since none of these existing taks-based libraries provided the flexibility
required to experiment with different scheduling and communication
techniques, (\swift is an interdisciplinary effort between
Computer Science and Astrophysics to study not only cosmological
phenomena, but also novel simulation algorithms and parallel computing techniques)
we chose to implement our own task scheduler
in \swift, which has since been back-ported as the general-purpose
\qs task scheduler \cite{gonnet2013quicksched}.
In \qs and \swift, task dependencies are specified explicitly,
as opposed to being implicitly derived from data dependencies,
allowing us to more easily build complex task hierarchies.
This also allowed us to extend the scheduler with the concept of
task conflicts and integrate the asynchronous communication
scheme described further on.

Despite its advantages, and the variety of implementations,
task-based parallelism is rarely used in
practice (notable exceptions include the PLASMA project
\cite{ref:Agullo2009} which uses QUARK/StarPU, and the {\tt deal.II} project
\cite{ref:Bangerth2007} which uses Intel's TBB).
The main problem is that to effectively use task-based parallelism,
most computations need to be completely redesigned to fit the paradigm,
which is usually not an option for large and complex codebases.

Since we were re-implementing \swift from scratch, this was not an issue.
The tree-based neighbour-finding described above was replaced with a more
task-friendly approach as described in \cite{ref:Gonnet2015}, in which
the domain is first decomposed into a grid of {\em cells} of edge length
larger or equal to the largest particle radius.
An initial set of interaction tasks is then defined over all cells and
pairs of neighbouring cells, such that if two particles are close enough to interact,
they are either in the same cell or they span a pair of neighbouring cells.
These initial interaction tasks are then refined by recursively
splitting cells that contain more than a certain number of particles
and replacing tasks that span a pair of split cells with tasks
spanning the neighboring sub-cells.
The resulting refined set of tasks contains all the cells and pairs of cells
over which particle interactions must be computed.

\begin{figure}
\centering
\includegraphics[width=\columnwidth]{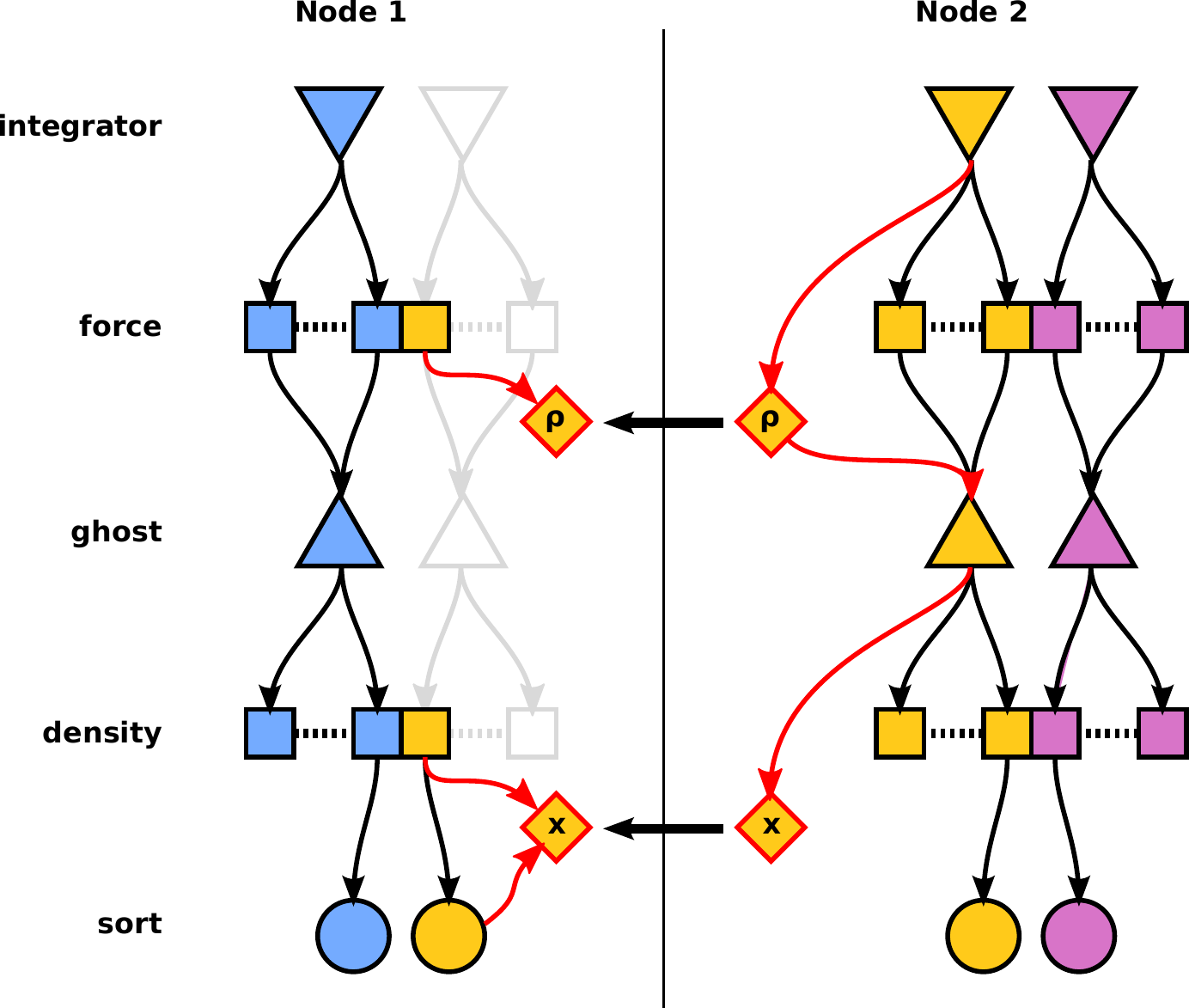}
\caption{Task hierarchy for the SPH computations in \swift,
  including communication tasks.
  Arrows indicate dependencies,
  i.e.~an arrow from task $A$ to task $B$ indicates that $A$
  depends on $B$. 
  The task color corresponds to the cell or
  cells it operates on, e.g.~the density and force tasks work
  on individual cells or pairs of cells.
  The task hierarchy is shown for three cells: blue, yellow,
  and purple. Although the data for the yellow cell resides on
  Node~2, it is required for some tasks on Node~1, and thus needs
  to be copied over  during
  the computation using {\tt send}/{\tt recv} tasks
  (diamond-shaped). \newline 
  Figure adapted from \cite{ref:Gonnet2015}.
  }
\label{tasks}
\end{figure}

The dependencies between the tasks are set following
equations \eqn{rho}, \eqn{dvdt}, and \eqn{dudt}, i.e.~such that for any cell,
all the tasks computing the particle densities therein must have
completed before the particle forces can be computed, and all the
force computations must have completed before the particle velocities
may be updated.
The task hierarchy is shown in Fig.~\ref{tasks}, where the particles in each
cell are first sorted (round tasks) before the particle densities
are computed (first layer of square tasks).
Ghost tasks (triangles) are used to ensure that all density computations
on a cell of particles have completed before the force evaluation tasks
(second layer of square tasks) execute.
Once all the force tasks on a cell of particles have completed,
the integrator tasks (inverted triangles) update the particle positions 
and velocities.
The decomposition was computed such that each cell contains $\sim 100$ particles,
which leads to tasks of up to a few milliseconds each.

Due to the cache-friendly nature of the task-based computations, 
and their ability to exploit symmetries in the particle interactions,
the task-based approach is already more efficient than the tree-based
neighbour search on a single core, and scales efficiently to all
cores of a shared-memory machine \cite{ref:Gonnet2015}.

\subsection{Task-based domain decomposition}

Given a task-based description of a computation, partitioning it over
a fixed number of {\em ranks} (using the MPI terminology)
is relatively straight-forward: we create
a {\em cell hypergraph} in which:
\begin{itemize}
  \item Each {\em node} represents a single cell of particles, and,
  \item each {\em edge} represents the tasks, connecting the
    cells.
\end{itemize}
Since in the particular case of \swift each task references at most
two cells, the cell hypergraph is just a regular {\em cell graph}.

Any partition of the cell graph represents a partition of the
computation, i.e.~the nodes belonging to each partition each belong
to a rank, and the
data belonging to each cell resides on the partition/rank to which
it has been assigned.
Any task spanning cells that belong to the same partition needs only
to be evaluated on that rank/partition, and tasks spanning more than
one partition need to be evaluated on both ranks/partitions.

\begin{figure}
\centering
\includegraphics[width=0.8\columnwidth]{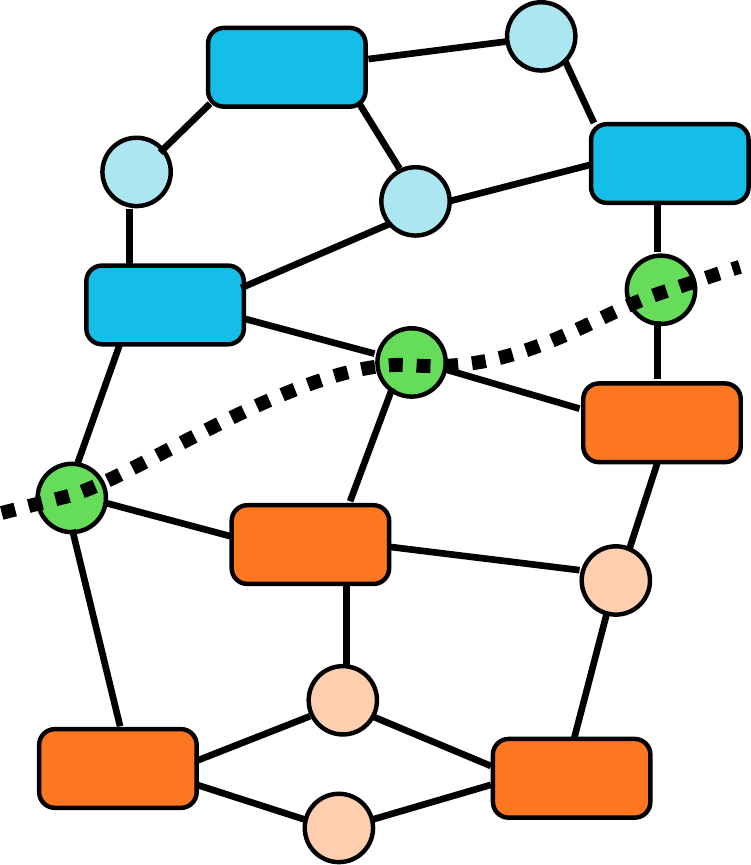}
\caption{Illustration of the task-based domain decomposition
  in which the tasks (circles) are {\em hyperedges} that connect one or
  more resources (rectangles). The resources are partitioned
  along the thick dotted line. The blue and orange tasks are
  executed on the respective partitions, whereas the green
  tasks/hyperedges along the cut line are executed on both.
  The cost of this partition is the sum of the green tasks,
  which are computed twice, as well as the cost imbalance
  of the tasks executed in each partition.}
\label{taskgraphcut}
\end{figure}

If we then weight each edge with the computational cost associated with
the tasks, then finding a {\em good} cell distribution reduces to finding a
partition of the cell graph such that the maximum sum of the weight
of all edges within and spanning in a partition is minimal
(see Fig.~\ref{taskgraphcut}).
Since the sum of the weights is directly proportional to the amount
of computation per rank/partition, minimizing the maximum sum
corresponds to minimizing the time spent on the slowest rank.
Computing such a partition is a standard graph problem and several
software libraries which provide good solutions\footnote{Computing
the optimal partition for more than two nodes is considered NP-hard.},
e.g.~METIS \cite{ref:Karypis1998} and Zoltan \cite{devine2002zoltan}, exist.

In \swift, the graph partitioning is computed using the METIS library.
The cost of each task is initially approximated via the
asymptotic cost of the task type and the number of particles involved.
After a task has been executed, it's effective computational cost
is computed and used.

Note that this approach does not explicitly consider any geometric
constraints, or strive to partition the {\em amount} of data equitably.
The only criteria is the computational cost of each partition, for
which the task decomposition provides a convenient model.
We are therefore partitioning the {\em computation}, as opposed
to just the {\em data}.

Note also that the proposed partitioning scheme takes neither the
task hierarchy, nor the size of the data that needs to be exchanged
between partitions/ranks into account.
This approach is therefore only reasonable in situations in which
the task hierarchy is wider than flat, i.e.~the length of the critical
path in the task graph is much smaller than the sum of all tasks,
and in which communication latencies are negligible.

\subsection{Asynchronous communications}

Although each particle cell resides on a specific rank, the particle
data will still need to be sent to any neighbouring ranks that have
tasks that depend on this data, e.g.~the the green tasks in
Fig.~\ref{taskgraphcut}.
This communication must happen twice at each time-step: once to send
the particle positions for the density computation, and then again
once the densities have been aggregated locally for the force
computation.

Most distributed-memory codes based on MPI \cite{ref:Snir1998}
separate computation and communication into distinct steps, i.e.~all
the ranks first exchange data, and only when the data exchange is
complete does computation start. Further data exchanges only happen
once computation has finished, and so on.
This approach, although conceptually simple and easy to implement,
has three major drawbacks:
\begin{itemize}
  \item The frequent synchronization points between communication
    and computation exacerbate load imbalances,
  \item the communication phase consists mainly of waiting on
    latencies, during which the node's CPUs usually run idle, and
  \item during the computation phase, the communication network
    is left completely unused, whereas during the communication
    phase, all ranks attempt to use it at the same time.
\end{itemize}

It is for these reasons that in \swift we opted for a fully
{\em dynamic and asynchronous} communication model in which local
data is sent whenever it is ready, data from other ranks is
only acted on once it has arrived, and there is no separation into
communication and computational phases.
In practice this means that no rank will sit idle waiting on
communication if there is any computation that can be done.

This fits in quite naturally within the task-based framework
by modelling communication as just another task type, i.e.~adding
tasks that send and receive particle data between ranks.
For every task that uses data that resides on a different rank,
{\tt send} and {\tt recv} tasks are generated automatically on the source
and destination ranks respectively.
At the destination, the task is made dependent of the {\tt recv}
task, i.e.~the task can only execute once the data has actually
been received.
This is illustrated in Fig.~\ref{tasks}, where data is exchanged across
two ranks for the density and force computations and the extra
dependencies are shown in red.

The communication itself is implemented using the non-blocking
{\tt MPI\_Isend} and {\tt MPI\_Irecv} primitives to initiate
communication, and {\tt MPI\_Test} to check if the communication was
successful and resolve the communication task's dependencies.  In the
task-based scheme, strictly local tasks which do not rely on
communication tasks are executed first.  As data from other ranks
arrive, the corresponding non-local tasks are unlocked and are
executed whenever a thread picks them up.

One direct consequence of this approach is that instead of a single
{\tt send}/{\tt recv} call between each pair of neighbouring ranks,
one such pair is generated for each particle cell.
This type of communication, i.e.~several small messages instead of
one large message, is usually strongly discouraged since the sum of
the latencies for the small messages is usually much larger than
the latency of the single large message.
This, however, is of no concern in \swift since nobody is actively
waiting to receive the messages in order, and the communication
latencies are covered by local computations.
A nice side-effect of this approach is that communication no longer
happens in bursts involving all the ranks at the same time, but
is more or less evenly spread over the entire computation, and is
therefore less demanding of the communication infrastructure.

%#####################################################################################################

\section{Scaling tests}

In this section we present some strong scaling tests of the \swift code on different
architectures for a representative cosmology problem.

\subsection{Simulation setup}

\begin{figure}
\centering
\includegraphics[width=\columnwidth]{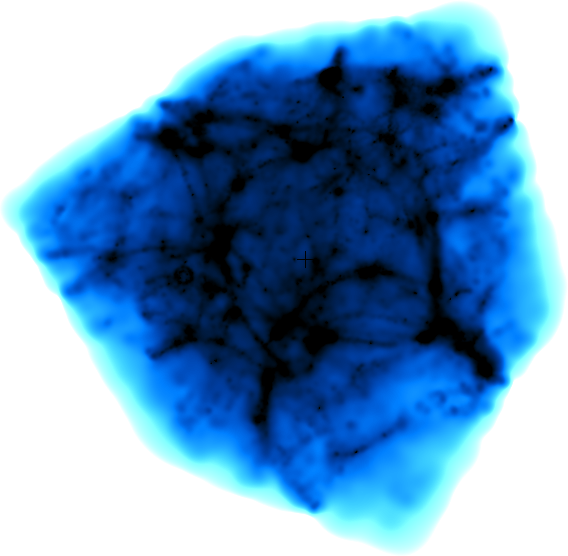}
\caption{The initial density field computed from the initial particle
  distribution used for our tests. The density $\rho_i$ of the particles spans 8
  orders of magnitude, requiring smoothing lengths $h_i$ changing by a factor of
  almost $1000$ across the simulation volume. \label{fig:ICs}}
\end{figure}

In order to provide a realistic setup, 
the initial particle distributions used in our tests were extracted by
resampling low-redshift outputs of the EAGLE project \cite{Schaye2015}, a
large suite of state-of-the-art cosmological simulations. By selecting outputs
at late times (redshift $z=0.5$), we constructed a simulation setup which is
representative of the most expensive part of these simulations, i.e.~when the
particles are highly-clustered and no longer uniformly distributed. This
distribution of particles is shown on Fig.~\ref{fig:ICs}.
In order to fit our simulation setup into the limited
memory of some of the systems tested, we have randomly down-sampled the particle
count of the output to $800^3=5.12\times10^8$, $600^3=2.16\times10^8$ and
$376^3=5.1\times10^7$ particles with periodic boundary conditions 
respectively. Scripts to generate these initial
conditions are provided with the source code. We then run the \swift code for
100 time-steps and average the wall clock time of these time-steps after having
removed the first and last ones, where disk I/O occurs.

The initial load balancing between nodes is computed in the first steps and
re-computed every 100 time steps, and is therefore not included in the timings.
Particles were exchanged between nodes whenever they strayed too far beyond
the cells in which they originally resided.

On all the machines, the code was compiled out of the box,
without any tuning, explicit vectorization, or exploiting any
other specific features of the underlying hardware. 

\subsection{x86 architecture: COSMA-5}

For our first test, we ran \swift on the COSMA-5 DiRAC2 Data Centric
System\footnote{\url{icc.dur.ac.uk/index.php?content=Computing/Cosma}}
located at the University of Durham. The system consists of 420 nodes
with 2 Intel Sandy Bridge-EP Xeon
E5-2670\footnote{\url{http://ark.intel.com/products/64595/Intel-Xeon-Processor-E5-2670-20M-Cache-2_60-GHz-8_00-GTs-Intel-QPI}}
CPUs clocked at $2.6~\rm{GHz}$ with each $128~\rm{GByte}$ of RAM. The
16-core nodes are connected using a Mellanox FDR10 Infiniband 2:1 blocking
configuration.

This system is similar to many Tier-2 systems available in most universities or
computing facilities. Demonstrating strong scaling on such a machine is
essential to show that the code can be efficiently used even on the type of
commodity hardware available to most researchers.

The code was compiled with the Intel compiler version \textsc{2016.0.1} and
linked to the Intel MPI library version \textsc{5.1.2.150} and METIS library
version \textsc{5.1.0}.

\begin{figure}
\centering
\includegraphics[width=\columnwidth]{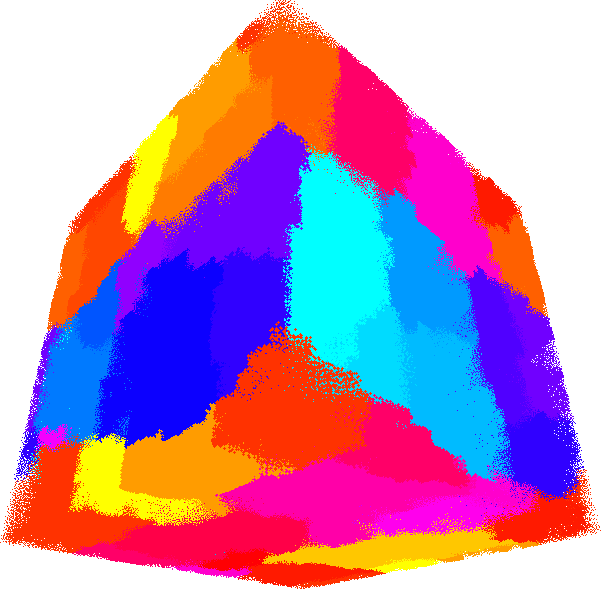}
\caption{The particles for the initial conditions shown on Fig.~\ref{fig:ICs}
  colored according to the node they belong to after a load-balancing call on
  32 nodes. As can be seen, the domain decomposition follows the cells in the mesh
  but is not made of regular cuts. Domains have different shapes and
  sizes. \label{fig:domains}}
\end{figure}

\begin{figure*}
\centering
\includegraphics[width=\textwidth]{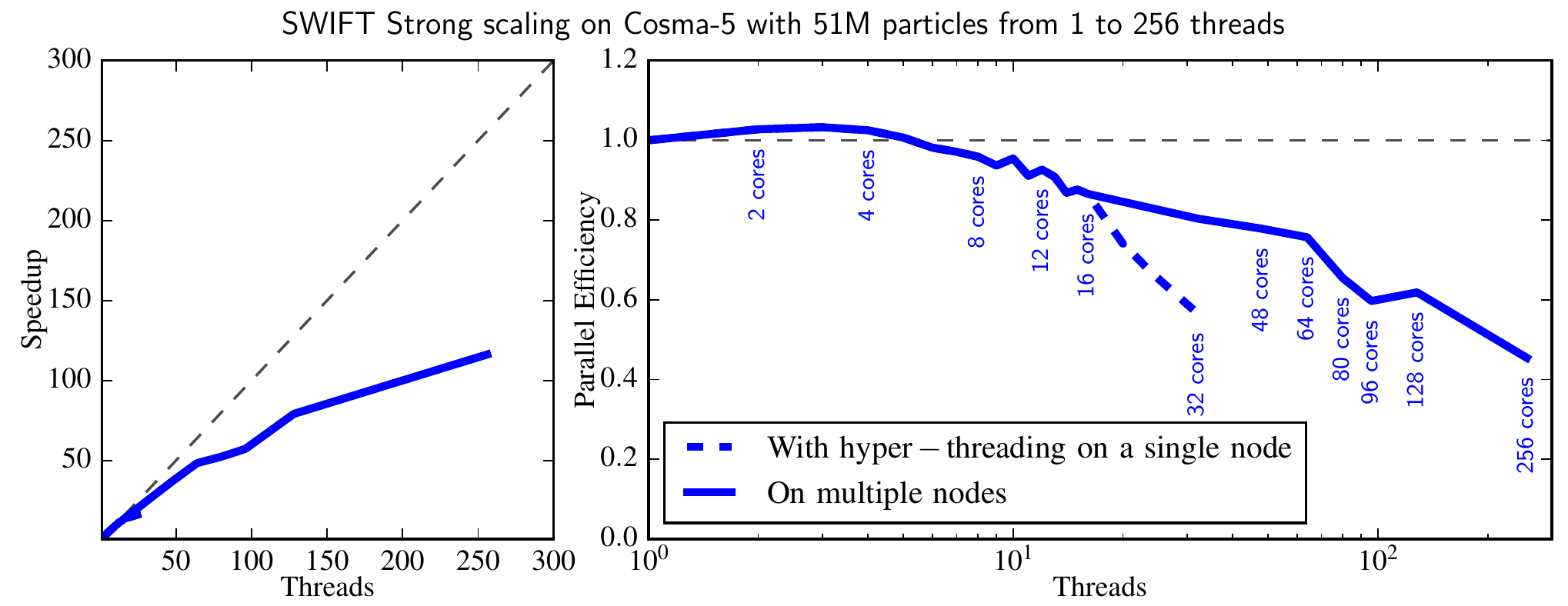}
\caption{Strong scaling test on the COSMA-5 machine (see text for hardware
  description). \textit{Left panel:} Code Speed-up. \textit{Right panel:}
  Corresponding parallel efficiency.  Using 16 threads per node (no use of
  hyper-threading) with one MPI rank per node, a good parallel efficiency (60\%)
  is achieved when increasing the thread count from 1 (1 node) to 128 (8 nodes)
  even on this relatively small test case. The dashed line indicates the
  efficiency when running on one single node but using all the physical and
  virtual cores (hyper-threading). As these CPUs only have one FPU per core, the
  benefit from hyper-threading is limited to a 20\% improvement when going
  from 16 cores to 32 hyperthreads.
  \label{fig:cosma}}
\end{figure*}

The simulation setup with $376^3$ particles was run on that system using 1 to
256 threads on 1 to 16 nodes, the results of which are shown on
Fig.~\ref{fig:cosma}. For this strong scaling test, we used one MPI rank per node and 16
threads per node (i.e.~one thread per physical core). We also ran on one single
node using up to 32 threads, i.e.~up to one thread per physical and
virtual core. Fig.~\ref{fig:domains} shows the domain decomposition
obtained via the task-graph decomposition algorithm described above for
the 16 node run.

\subsection{x86 architecture: SuperMUC}

For our next test, we ran \swift on the SuperMUC x86 phase~1 thin
nodes \footnote{\url{https://www.lrz.de/services/compute/supermuc/systemdescription/}}
located at the Leibniz Supercomputing Center in Garching near Munich,
currently ranked 23rd in the Top500 list\footnote{\url{http://www.top500.org/list/2015/11/}}. This
system consists of 9\,216 nodes with 2 Intel Sandy Bridge-EP Xeon E5-2680
8C\footnote{\url{http://ark.intel.com/products/64583/Intel-Xeon-Processor-E5-2680-(20M-Cache-2_70-GHz-8_00-GTs-Intel-QPI)}}
at $2.7~\rm{GHz}$ CPUS. Each 16-core node has $32~\rm{GByte}$ of RAM.
The nodes are split in 18
``islands'' of 512 nodes within which communications are handled via an
Infiniband FDR10 non-blocking Tree. These islands are then connected using a 4:1
Pruned Tree.

This system is similar in nature to the COSMA-5 system used in the previous set
of tests but is much larger, allowing us to test scalability at a much larger
scale.

The code was compiled with the Intel compiler version \textsc{2015.5.223} and
linked to the Intel MPI library version \textsc{5.1.2.150} and METIS library
version \textsc{5.0.2}.

\begin{figure*}
\centering
\includegraphics[width=\textwidth]{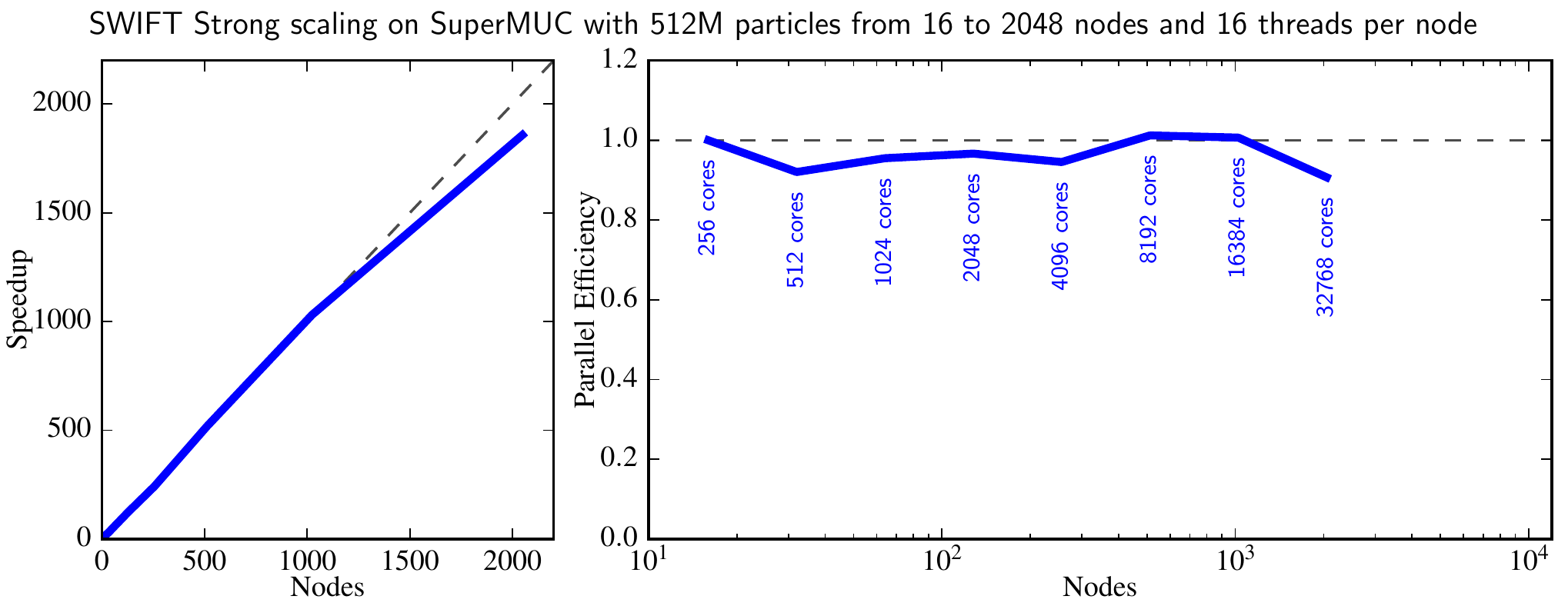}
\caption{Strong scaling test on the SuperMUC phase~1 machine (see text
  for hardware description). \textit{Left panel:} Code
  Speed-up. \textit{Right panel:} Corresponding parallel efficiency.
  Using 16 threads per node (no use of hyper-threading) with one MPI rank
  per node, an almost perfect parallel efficiency is achieved when
  increasing the node count from 16 (256 cores) to 2\,048 (32\,768
  cores).
  \label{fig:superMUC}}
\end{figure*}

The simulation setup with $800^3$ particles was run using 16 to
2048 nodes (4 islands) and the results of this strong scaling test are shown in
Fig.~\ref{fig:superMUC}. For this test, we used one MPI rank per node and 16
threads per node, i.e.~one thread per physical core.

\subsection{Blue Gene architecture: JUQUEEN}

For our last set of tests, we ran \swift on the JUQUEEN IBM Blue Gene/Q
system\footnote{\url{http://www.fz-juelich.de/ias/jsc/EN/Expertise/Supercomputers/JUQUEEN/Configuration/Configuration_node.html}}
located at the J\"ulich Supercomputing Center,
currently ranked 11th in the Top500 list.
This system consists of
28\,672 nodes with an IBM PowerPC A2 processor running at
$1.6~\rm{GHz}$ and $16~\rm{GByte}$ of RAM each. Of notable interest
is the presence of two floating units per compute core. The system is
composed of 28 racks containing each 1\,024 nodes. The network uses a
5D torus to link all the racks.

This system is larger than the SuperMUC supercomputer described above and
uses a completely different processor and instruction set.
We use it here to demonstrate that our results are not dependant
on the hardware being used.

The code was compiled with the IBM XL compiler version \textsc{30.73.0.13} and
linked to the corresponding MPI and METIS library
versions \textsc{4.0.2}.

\begin{figure}
\centering
\includegraphics[width=\columnwidth]{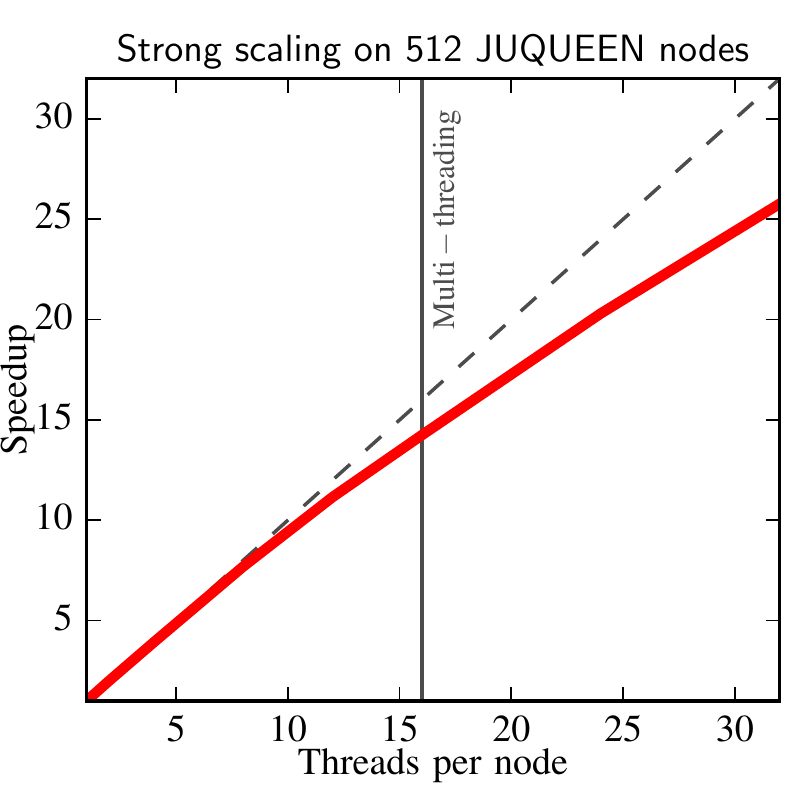}
\caption{Strong scaling test of the hybrid component of the code. The
  same calculation is performed on 512 node of the JUQUEEN Blue Gene
  supercomputer (see text for hardware description) using a single MPI
  rank per node and varying only the number of
  threads per node. The code displays excellent scaling even when all the cores and
  hardware multi-threads are in use. \label{fig:JUQUEEN1}}
\end{figure}

The simulation setup with $600^3$ particles was first run using
512 nodes with one MPI rank per node and varying only the number of threads per
node. The results of this test are shown in Fig.~\ref{fig:JUQUEEN1}.

\begin{figure*}
\centering
\includegraphics[width=\textwidth]{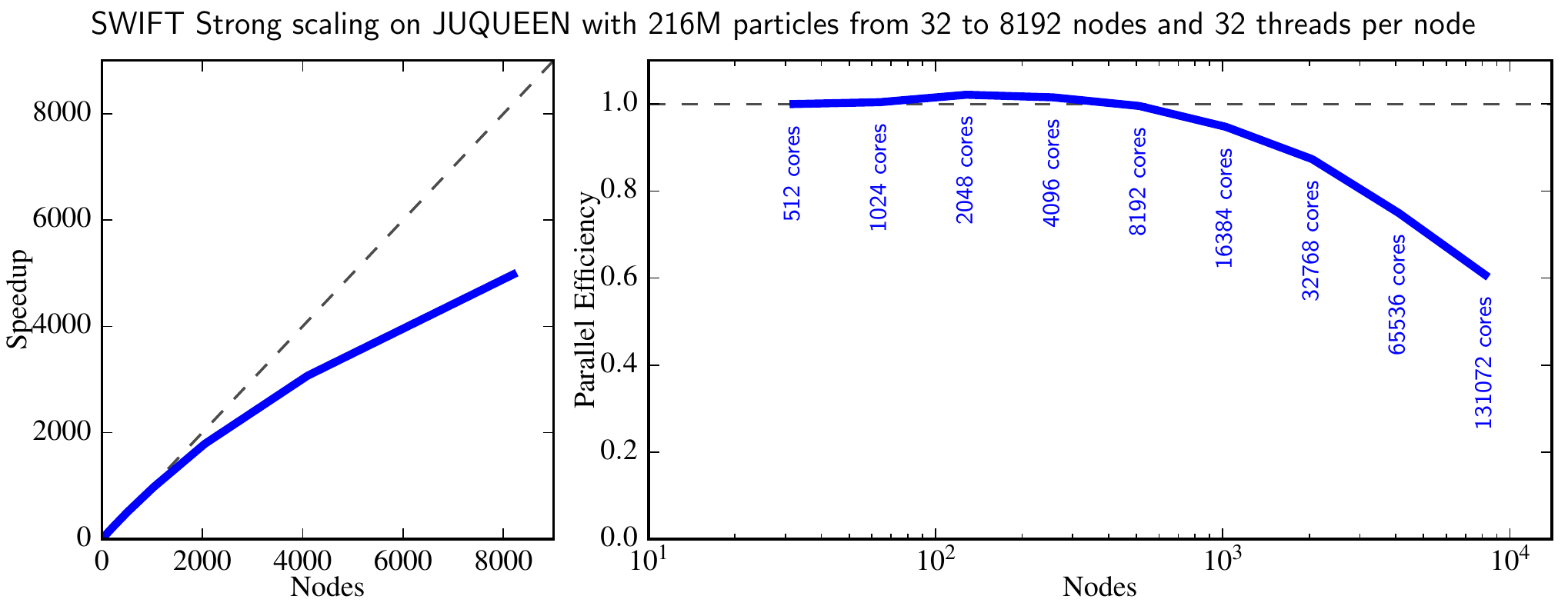}
\caption{Strong scaling test on the JUQUEEN Blue Gene machine (see text
  for hardware description). \textit{Left panel:} Code
  Speed-up. \textit{Right panel:} Corresponding parallel efficiency.
  Using 32 threads per node (2 per physical core) with one MPI rank
  per node, a parallel efficiency of more than $60\%$ is achieved when
  increasing the node count from 32 (512 cores) to 8\,192 (131\,072
  cores). On 8\,192 nodes there are fewer than 27\,000 particles per
  node and only a few hundred tasks, making the whole problem
  extremely hard to load-balance effectively.
  \label{fig:JUQUEEN2}}
\end{figure*}

We later repeated the test, this time varying the number of nodes from 32 to
8192 (8 racks).  For this test, we used one MPI rank per node and 32 threads per
node, i.e.~two threads per physical core. The results of this strong scaling
test are shown in Fig.~\ref{fig:JUQUEEN2}.

%#####################################################################################################

\section{Discussion \& conclusions}

The strong scaling results presented in the previous sections on
three different machines demonstrate the ability of our framework
to scale on both small commodity
machines, thanks to the use of task-based parallelism at the node level, and on
the largest machines (Tier-0 systems) currently available, thanks to the
task-based domain distribution and asynchronous communication schemes.
We would like to emphasize that these results were obtained for a
realistic test case without any micro-level optimization or explicit
vectorization.

Excellent strong scaling is also achieved when increasing the number of threads
per node (i.e.~per MPI rank, see fig.~\ref{fig:JUQUEEN1}), demonstrating that
the description of MPI (asynchronous) communications as tasks within our
framework is not a bottleneck. One common conception in HPC is that the number
of MPI communications between nodes should be kept to a minimum to optimize the
efficiency of the calculation. Our approach does exactly the opposite with large
number of point-to-point communications between pairs of nodes occurring over the
course of a time-step. For example, on the SuperMUC machine with 32 nodes (512
cores), each MPI rank contains approximately $1.6\times10^7$ particles in
$2.5\times10^5$ cells. \swift will generate around $58\,000$ point-to-point
asynchronous MPI communications (a pair of \texttt{send} and \texttt{recv} tasks)
{\em per node} and {\em per timestep}. Each one of these communications involves,
on average, no more than 6\,kB of data. Such an insane number of small messages is
discouraged by most practitioners, but seems to works well in practice.
Dispatching communications
over the course of the calculation and not in short bursts, as is commonly done,
may also help lower the load on the network.

One time-step on $8\,192$ nodes of the JUQUEEN machine takes $63~\rm{ms}$ of
wall-clock time. All the loading of the tasks, communications and running of the
tasks takes place in that short amount of time. Our framework can therefore
load-balance a calculation over $2.6\times10^5$ threads with remarkable
efficiency.

We emphasize, as was previously demonstrated in \cite{ref:Gonnet2015}, that \swift
is also much faster than the \gadget code \cite{Springel2005}, the
\emph{de-facto} standard in the field of particle-based cosmological
simulations. The simulation setup that was run on the COSMA-5
system takes $2.9~\rm{s}$ of wall-clock time per time-step on $256$ cores using
\swift whilst the default \gadget code on exactly the same setup with the same
number of cores requires $32~\rm{s}$.
The excellent scaling
performance of \swift allows us to push this number further by simply increasing
the number of cores, whilst \gadget reaches its peak speed (for this problem) at
around 300 cores and stops scaling beyond that. This unprecedented scaling
ability combined with future work on vectorization of the calculations within
each task will hopefully make \swift an important tool for future simulations in
cosmology and help push the entire field to a new level.

These results, which are in no way restricted to astrophysical simulations,
provide a compelling argument for moving away from the traditional
branch-and-bound paradigm of both shared and distributed memory programming
using synchronous MPI and OpenMP.
Although fully asynchronous methods, due to their somewhat anarchic nature,
may seem more difficult to control, they are conceptually simple and easy
to implement\footnote{The \swift source code consists of less than 21\,K lines
of code, of which roughly only one tenth are needed to implement the task-based
parallelism and asynchronous communication.}.
The real cost of using a task-based approach comes from having
to rethink the entire computation to fit the task-based setting. This may,
as in our specific case, lead to completely rethinking the underlying
algorithms and reimplementing a code from scratch.
Given our results to date, however, we do not believe we will regret this
investment.

\swift, its documentation, and the test cases presented in this paper are all
available at the address \web.

%#####################################################################################################

\section{Acknowledgements}
This work would not have been possible without Lydia Heck's help and
expertise. We acknowledge the help of Tom Theuns, James Willis, Bert
Vandenbroucke, Angus Lepper and other contributors to the \swift
code. \\ We thank Heinrich Bockhorst and Stephen Blair-Chappell from {\sc
  intel} as well as Dirk Brommel from the J\"ulich Computing Center
and Nikolay J. Hammer from the Leibniz Rechnenzentrum for their help
at various stages of this project.\\ This work used the DiRAC Data
Centric system at Durham University, operated by the Institute for
Computational Cosmology on behalf of the STFC DiRAC HPC Facility
(\url{www.dirac.ac.uk}). This equipment was funded by BIS National
E-infrastructure capital grant ST/K00042X/1, STFC capital grant
ST/H008519/1, and STFC DiRAC Operations grant ST/K003267/1 and Durham
University. DiRAC is part of the National E-Infrastructure. This work
was supported by the Science and Technology Facilities Council
ST/F001166/1 and the European Research Council under the European
Union's ERC Grant agreements 267291 ``Cosmiway'', and by {\sc intel}
through establishment of the ICC as an {\sc intel} parallel computing
centre (IPCC).

\bibliographystyle{abbrv}
\bibliography{biblio}

\end{document}